\documentclass[times, 10pt,twocolumn]{article}
\usepackage{latex8}
\usepackage{epsfig}
\usepackage{epic}

\begin{document}
\author{P. Oscar Boykin\\
Department of Electrical and Computer Engineering\\
University of Florida\\
Gainesville, FL, USA\\
boykin@ece.ufl.edu\\
\and
Vwani P. Roychowdhury\\
Electrical Engineering Department\\
University of California, Los Angeles\\
Los Angeles, CA, USA\\
vwani@ee.ucla.edu\\
}

\title{Reversible Fault-Tolerant Logic}
\maketitle

\begin{abstract}
It is now widely accepted that the CMOS technology implementing irreversible 
logic will hit a scaling limit beyond 2016, and that the increased power
dissipation is a major limiting factor. Reversible computing can potentially
require arbitrarily small amounts of energy.
Recently several nano-scale devices 
which have the potential to scale, and which naturally perform reversible
logic, have emerged.
This paper addresses several fundamental issues that need to be
addressed
 before any nano-scale reversible computing systems can be realized, including
 reliability and performance trade-offs and architecture optimization.  Many
 nano-scale devices will be limited to only near neighbor interactions, 
requiring careful optimization of circuits. We provide efficient fault-tolerant
(FT) circuits
when restricted to both 2D and 1D.  Finally, we compute bounds on the 
entropy 
(and hence, heat) generated by our FT circuits and provide quantitative
estimates on how large can we make our circuits before we lose any advantage
over irreversible computing.
\end{abstract}

\section{Introduction}
\label{sec:intro}
There are several compelling reasons for a renewed interest in reversible
computing systems: First, it is now widely accepted that the CMOS technology
implementing irreversible logic will hit a scaling limit beyond 2016, and the
increased power dissipation is a major limiting factor. Reversible
computing\cite{Bennett73,Toffoli80,FT82} can potentially require zero or very little
energy. Second, several new nano-scale devices 
which have the
potential to scale, and which naturally perform reversible logic, have emerged.
This paper
addresses several fundamental issues that need to be addressed before any
nano-scale reversible computing systems can be realized, including:
\begin{enumerate}
\item {\em Reliability and Performance Trade-offs:}
Current nano-scale logic
proposals appear to provide extremely unreliable devices,
requiring extensive use of
fault-tolerant (FT) circuits. We provide a systematic design for reversible
FT circuits, which will work reliably {\em even if each gate
has an error probability as high as $1/108$}. We also calculate the blow up in
size and gate count that would result from the use of such FT circuits.
\item {\em Architecture Optimization:}
many of the proposed nano-scale devices will be limited to only near-neighbor
interactions\cite{Lloyd93,Kane98,VYWJBRMD00}, requiring careful circuit
optimization. We provide efficient FT circuits when restricted to both 2D and
1D. We show that for a 2D topology with near neighbor connections, the error
threshold decreases only to  $1/273$, and that a 1D lattice that is 27 bits
wide but arbitrarily long has an error threshold only $23\%$ less
than the full 2D case.
\item {\em Power Dissipation:} Reversible computing systems in the presence of errors will generate heat. We compute bounds on the entropy (and hence, heat) generated by our FT circuits and provide quantitative estimates on how large can we make our circuits before we lose any advantage over irreversible computing. 
\end{enumerate}

Many previous works have considered gate-level fault tolerance techniques for
irreversible gates\cite{vonneumann56,NSF02,HJ02,GQF04}.  Local fault tolerance schemes for irreversible
automata have also been studied \cite{Gacs86}.
Quantum computers are reversible; however, the properties of
quantum errors and quantum information are sufficiently different
from the classical case
that fault-tolerant quantum computation\cite{Shor96,Preskill98}
is not directly applicable to the
traditional reversible classical computing model, which is the subject
of this work.

This paper is organized into three main parts.
In Section \ref{sec:revmp}, we give our model of noisy reversible gates and
give a circuit fault-tolerant error-recovery and compute the overhead that the
circuit requires both in time and space.  We show that, as long as the error
rate is below a threshold, the circuit can be made reliable.
In Section \ref{sec:localft}, we
apply the results of Section \ref{sec:revmp} to locally connected models
where bits can only operate on their nearest neighbors.  We consider the local
problem in $1D$ and $2D$.  In Section \ref{sec:entropy}, we compute how much
entropy and heat is dissipated in the error-recovery process, and we see that
the entropy saving aspect of reversible computing is lost
in our scheme once the error rate gets close to the threshold.

\section{Reversible majority multiplexing}
\label{sec:revmp}
\begin{figure}
\centering
\includegraphics{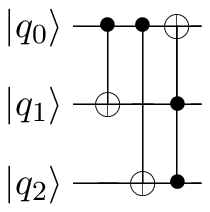}
\caption{The reversible Majority gate constructed
from two controlled-not gates and one Toffoli gate.
The horizontal lines represent bits.
Each vertical line connecting
horizontal lines represents a gate.  If every bit
connected by a filled dark circle on a particular vertical
line has the value one,
the bit on the horizontal line with the $\oplus$ is flipped.  Time flows from
left to right.}
\label{fig:maj}
\end{figure}

In standard irreversible computing, we often imagine that functions are
represented by circuits of wires and gates at fixed positions.  We can think
of the bits as moving through the wires to different gates.  In reversible
computing, since gates have identical numbers of input bits and output bits,
we have a choice:
we can picture the bits moving through wires to gates at fixed locations,
or picture the bits as fixed locations in space and
the reversible function as a sequence of reversible gates
\emph{applied on the bits}.  This is commonly represented as the gate array
notation where space is on the y-axis and time is on the x-axis, and
operations are boxes or symbols that connect the bits they are applied to
(see, for
instance Figure \ref{fig:maj}).  This model is
realizable in many nanocomputing proposals\cite{Lloyd93,Kane98,VYWJBRMD00}\footnote{It should be
noted that, since all quantum computers are reversible, this work is
particularly applicable to classical processing on quantum hardware.}

The error model is a simple independent gate failure model: at each
application, a gate will randomize all the bits it is
applied to with probability $g$.  While this model does not directly address correlated failures,
it will apply as long as the probability that $k$ out of $G$
gates fail is \emph{less than} ${G \choose k}g^k(1-g)^{G-k}$.
For most of this paper, we will assume that this error rate
applies to any three-bit gate, and that we have access to no fault-free
operations.  Our goal is to make larger modules of $T$ reversible gates with
a module error rate which is independent of $T$, and is in general much smaller
than $1-(1-g)^T\approx gT$ (which is what we could expect without any fault
tolerance).
In fact, we show that by using $O(T\log^{4.75} T)$ gates instead of just $T$,
we obtain an error rate that is constant in $T$.

Rather than explicitly deal with error correction codes, the best gate-level,
fault-tolerant schemes for classical computing are those based on Von-Neumann
multiplexing\cite{vonneumann56}.  In this case, each bit is copied to a second bit, a random
permutation is applied, and finally, a NAND gate is applied to each pair of
bits.  These approaches make use of the irreversible NAND gate.  Schemes
such as this can result in fault-tolerant computation as long as the gate
error rate is less than about $11\%$\cite{vonneumann56}.

Rather than base our multiplexing scheme on NAND or some derivative, we base
ours on a reversible extension of the majority gate (MAJ), depicted in terms of
CNOT and Toffoli in Figure \ref{fig:maj}\footnote{We note that variants
of the MAJ gate
have found application in algorithmic cooling\cite{FLMR04} and reversible
addition\cite{CDKM04}; thus MAJ appears to be a valuable gate for reversible
and quantum computers.}.
The reversible majority gate (MAJ), has a truth table given in Table
\ref{tab:maj_truth}.

\begin{table}
\centering
\begin{tabular}{r|l}
Input & Output \\
\hline\\
000 & 000\\
001 & 001\\
010 & 010\\
011 & 111\\
100 & 011\\
101 & 110\\
110 & 101\\
111 & 100\\
\end{tabular}
\caption{The truth table for the reversible MAJ gate.  Note that each input
has a unique output, and that the first bit of the output is the
majority of the input bits.  This gate is obtained by flipping the second two
bits if the first bit is 1, and then flipping the first bit if the second two
bits are 1.}
\label{tab:maj_truth}
\end{table}

\begin{figure}
\centering
\scalebox{0.8} {
\includegraphics{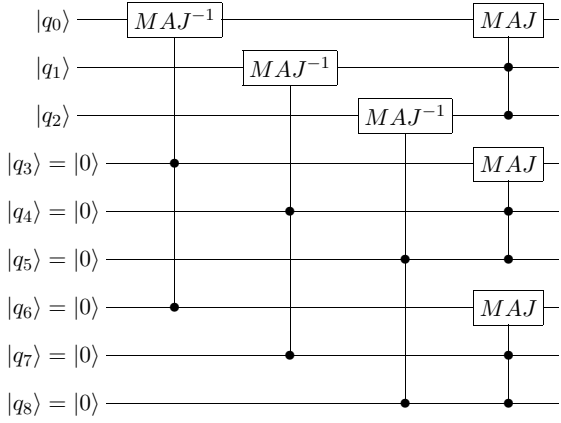}
}
\caption{A reversible multiplexing scheme based on the 3-bit repetition
code.  The output bits are those bit positions 0,3,6.  The rest may be
discarded.  This circuit is the error-recovery circuit referred to
as $E_L$ in Figure \ref{fig:concat}}
\label{fig:circuit_opt}
\end{figure}

We claim that the circuit in Figure \ref{fig:circuit_opt} is in fact a
fault-tolerant error correction circuit.  To see this, consider the code space of
$000$ representing logical zero ($0_L$) and $111$ ($1_L$) representing logical one.
Consider
Figure \ref{fig:circuit_opt} as having two phases, encoding and decoding.  The
first three $MAJ^{-1}$ gates are the encoding gates and the last three $MAJ$
gates are decoding gates.  After the encoding gates the bits should
all have the same value (i.e. $000000000$ or $111111111$ if the input was
$0_L$ or $1_L$, respectively).  Decoding puts the majority value of each block
of three bits into the three output bits.  Clearly, if there are no errors, the
circuit should output exactly what was input, due to the
symmetry of the code under permutations.  Now we simply observe that, if any
single error occurs, it will change at most one bit in each of the final
decoder blocks.  Since the decoders return the majority result in their output
bit, a single bit flip will not change the majority result.  If one of the
final MAJ gates has an error, it will only effect one bit in the output, and
that can be repaired in the next error-recovery cycle.  Thus we have a
fault-tolerant error-recovery because we can tolerate errors in our recovery
gates.

Since the codewords in this system are repetition code words, we can use any
universal, reversible set of gates for computation directly on the repetition
codewords.  After each gate operation, we apply our error-recovery circuit
from Figure \ref{fig:circuit_opt}.
Now that we have a circuit, we can ask: for what error rates will this circuit
perform as expected?  

\subsection{Concatenation}
\label{sec:concat}
In order to suppress the probability of error, we code one bit as three bits
in a repetition code.  But why stop there?  We could also code each of the
three bits into another repetition code, resulting in nine bits, and so forth.
This method is called concatenation.

We say a physical bit is level $0$, a three-bit code $000$ or $111$ is level
$1$, a nine bit code is level $2$, a $3^L$ bit code is level $L$.
A bit encoded at level $L$ is made up of three bits encoded at level
$L-1$.  A gate at level $0$ is a physical gate to which we have
access (in our case, $MAJ$).
To implement a logical gate at level $1$, we apply the gate at
level $0$ to each of the three bits in the code for level $1$.
To implement a 3-bit gate at level $L$,
we apply the gate at level $L-1$ on the logical bits in the code,
and then correct any errors we may have caused in each
of the bits.  This is depicted in Figure \ref{fig:concat}.

\subsection{The threshold for fault-tolerant computation}
\label{sec:threshold}
Throughout this work, we consider the gate error rate $g$ as the error
rate for a 3-bit operation.  Thus, we assume that we can reset three bits with
one initialization operation.
The error-recovery circuit depicted in Figure \ref{fig:circuit_opt}
requires us to initialize six bits (two 3-bit initialization operations),
apply three $MAJ^{-1}$ gates, and three $MAJ$ gates for a total of
eight gate operations
(six if initialization can be assumed to be far more accurate than our gates).  As
previously shown, as long as there is no more than one error in all
of these operations, the final result will not be an error.  We say that the
error-recovery circuit requires $E$ gates.

In addition to the error-recovery, we also have the logical gate which we
want to apply on the data.  To apply our logical gate, we need to
operate on each of the three bits in the code, which gives us three more gates
which can go wrong.

A particular bit will be correct unless there are two or more errors.
Thus, if each
operation has an error rate $g$ and there are $G=3+E$ operations acting
on each encoded bit (some act on more than one encoded bit),
the bit error rate $P_{bit}$ is:
\begin{eqnarray*}
P_{bit} &\le& \sum_{k=2}^{G}{G \choose k} g^k(1-g)^{G-k}\\
              &\le& {G \choose 2}g^2
\end{eqnarray*}

The probability that a gate has no error is at least as large
as the probability that none of the bits are in error if each were considered
independently:
\[
1-g_{logical} \ge (1 - P_{bit})^3
\]
The above is true because the right hand side triple counts the case where the
logical gate (the gate applied to all three bits) fails.
We note that the above bound is a convenient bound, but a tighter bound will result
in an improved error threshold.
We can use the above
to see that:
\begin{eqnarray}
g_{logical} &\le& 1 - (1 - P_{bit})^3\nonumber\\
            &\le & 3 P_{bit}\nonumber\\
	    &\le & 3{G \choose 2}g^2\label{eq:rem_err}
\end{eqnarray}

Thus if we want $g_{logical} < g$, it is sufficient to have
$g < \frac{1}{3{G \choose 2}}$,
which we call the threshold.  In our cases of $G=11$ (3 + $(E=8)$) and
$G=9$ (3 + $(E=6)$), we get
threshold results of $\rho=1/165$ and $\rho=1/108$, respectively.

The above only shows that we can decrease the logical error rate by applying
the error-recovery circuit; it does not show that we can make it as small as
we like.  In order to push the error rate to lower values, we
\emph{concatenate} our bits recursively.  At the lowest logical level, $L=0$,
each bit is represented by a physical bit.  At all other levels, each bit at
level $L$ is represented by 3-bits at level $L-1$.  Thus, at level $L$ we are
actually using $3^L$ physical bits.  We only pay attention to the error rates
at the largest level, but after each gate at level $L$ we do an error-recovery
at level $L$, which in turn applies gates at level $L-1$, and so on, until we
reach the bottom level.  This recursive structure is represented in figure
\ref{fig:concat}.

Thus, if the error
probability at level $k$ is $g_k$, $g_0$ = $g$, and we have $G$ total operations
to perform, Equation \ref{eq:rem_err} tells us that:
\begin{eqnarray*}
g_{k+1} &\le& 3{G\choose 2} g_k^2
\end{eqnarray*}
To solve equations like the one above, we introduce $\kappa = \log 3{G\choose
2}$, and $r_k = \log g_k$:
\begin{eqnarray}
g_{k+1} &\le& 3{G\choose 2} g_k^2 \nonumber\\
\log g_{k+1} &\le& \log 3{G\choose 2} + 2\log g_k\nonumber\\
r_{k+1} &\le& \kappa + 2 r_k\nonumber\\
(r_{k+1} + \kappa) &\le& 2(r_k + \kappa)\nonumber\\
r_k &\le& 2^k(r_1 + \kappa) - \kappa\nonumber\\
g_k &\le& \frac{(3{G \choose 2} g_0)^{2^k}}{3{G \choose 2}}\nonumber\\
    &=& \rho \left(\frac{g}{\rho}\right)^{2^k}\label{eq:g_k}
\end{eqnarray}
Assuming $G=9$, $3{G \choose 2}=108$; thus, if the lowest level gate
error is $g_0 < 1/108$, we can correct all the errors with high probability if
we use enough levels of concatenation.
Throughout this paper we use $\rho$ for threshold values.   Thus, if
we define $\rho = 1/3{G \choose 2}$, and 
if $g_0 < \rho$ we are sure to be able to correct all errors in the limit of
large $L$.


\subsection{Circuit blowup}
\label{sec:circuit_blowup}
\begin{figure}
\scalebox{0.7}{
\input{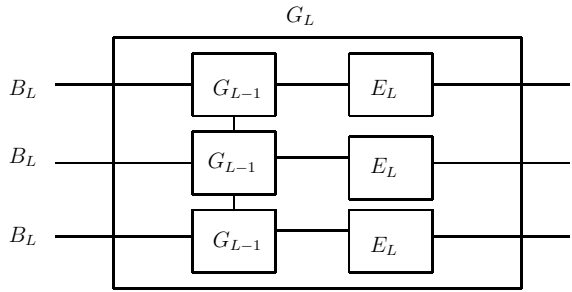}
}
\caption{A gate at concatenation level $L$.  We apply
the gate three times at level $L-1$ and then do an error correction cycle.
$G_{L-1}$ is the gate operation at the lower logical level, $E_L$
is the error-recovery circuit (Figure \ref{fig:circuit_opt})
on logical level $L$ (using only gates at level $L-1$).
}
\label{fig:concat}
\end{figure}
\label{sec:blowup}
In this section, we consider how much larger a module made of $T$ perfect gates
will be when constructed from the FT techniques of this paper, such that the
final module will have a constant probability of error irrespective of T.

To implement a 3-bit gate at level $L$, we must correct 3-bits
at level $L-1$.
Starting from Figure \ref{fig:concat}, we can see that
if $\Gamma_k$ is the number of gates required for one complete error correction
and gate operation
on levels $k$ and lower, and $E$ is the number of gates
required to do error-recovery, then we have that:
\begin{eqnarray*}
\Gamma_k &=& 3\Gamma_{k-1} + 3 E \Gamma_{k-1}\\
         &=& 3(1+E)\Gamma_{k-1}\\
         &=& \left( 3(1+E)\right)^k
\end{eqnarray*}
Using $G=3+E$, we have:
\[
\Gamma_k = \left( 3(G-2)\right)^k
\]
So there is an exponential blow-up in gate count with concatenation depth.
The bit size blowup is just as easy to calculate: at each level, we use 9 bits
of the level below:
\begin{eqnarray*}
S_{k+1} &=& 9 S_k\\
S_k &=& 9^k
\end{eqnarray*}

How deep do we need to concatenate?
If the module we want to simulate has $T$ gates, we need $g_L \le 1/T$
in order to have less than on average one error in the FT module. Hence,
by bounding Equation \ref{eq:g_k}:
\begin{eqnarray}
\rho (g/\rho)^{2^L} &\le& \frac{1}{T}\nonumber\\
L &\ge& \log_2 \frac{ \log T \rho }{\log \rho/g}\label{eq:level_bound} 
\end{eqnarray}
If we use the minimum valid value for $L$,
the gate blow-up factor is $\Gamma_L = O(G^L)$, and is poly-log in $T$:
\begin{eqnarray*}
(3(G-2))^L &=& 2^{L \log_2 3(G-2) }\\
    &=&\left(\frac{\log T \rho}{\log \rho/g}\right) ^ {\log_2
    3(G-2)}
\end{eqnarray*}
As is the size blow-up factor:
\begin{eqnarray*}
S_L=9^L&=& 2^{L \log_2 9}\\
   &=& \left(\frac{\log T \rho}{\log \rho/g}\right) ^ {\log_2 9}\\
   &\approx& \left(\frac{\log T \rho}{\log \rho/g}\right) ^ {3.17}
\end{eqnarray*}
For $G=11$, we have $(3(G-2))^L=O((\log T)^{4.75})$ and $S_L=O((\log
T)^{3.17})$.

Suppose we have $g=\rho/10$ with $G=9$ and $\rho\approx 10^{-2}$.  Without
any error correction, modules larger than $1,000$ gates will almost certainly
be faulty.
If we want to make a module of $T=10^6$,
we need $L = \log_2 ((\log_2 10^4)/\log_2 10)=2$, so we can make an
accurate module with $10^6$ gates, using 2 levels of concatenation, if
$g=\rho/10$.  This means that, rather than using one gate, we will need to replace
each with $(3(G-2))^2= 441$ gates and replace each bit with $3^2=81$ bits.
However, we are able to construct a much
larger module: $10^6$ logical gates rather than $1,000$ logical gates.

We have shown that we can take noisy gates and create modules of bounded
noise with only a poly-log overhead factor.  Once we have modules with bounded
noise, higher level fault tolerance techniques may be applied.

\section{Local reversible fault-tolerant schemes}
\label{sec:localft}
In this section, we consider the problem of an array of bits on which we may
operate with noisy, reversible gates.  We assume that we may only operate on at
most three neighboring bits at a time.  When it is necessary to operate on
pairs of remote bits, we must first move them close together by a series of
SWAP operations and then operate.  This introduces extra overhead into every
logical operation.  Since we are assuming that each operation is noisy we
expect this to reduce the error threshold.  We will first consider a 2D array
and see that the 2D array only requires extra SWAP operations to operate on three
\emph{logical} bits, and no extra SWAP operations to do error-recovery.
As such, the threshold is not much lower than the result of Section
\ref{sec:threshold}.  Later, we
consider a 1D array; in this case, our error correction circuit will require
many SWAP operations.  We can expect this to lower the threshold much more
than the 2D case.

\subsection{A local 2-dimensional fault-tolerant system}
\label{sec:2dft}
In Figure \ref{fig:circuit_opt}, we presented a fault-tolerant
error correction circuit which assumes that any pair of bits may be
operated on; notice, however, that during the recovery process,
only certain bits interacted.
In practice, many systems may allow only local
operations.  In this section we consider a 2D lattice of bits.  At
each time step any adjacent pairs (or triples) of bits may interact.

\begin{figure*}
\centering
\scalebox{0.75}{
\includegraphics{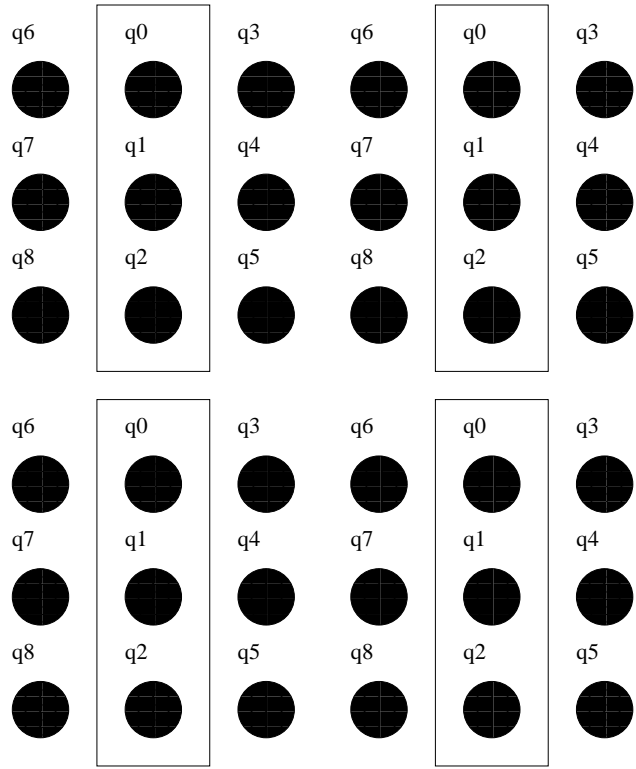}
}
\caption{Layout of bits on a 2D lattice.  Boxes denote the locations of
the logical bits; the other bits are ancillary bits.  To bring logical
bits close to one another, we can either move them perpendicular to the
logical line (q0,q1,q2) by swapping them past the ancillary bits in
between two logical lines (q3,q4,q5 for one bit and q6,q7,q8 for the other),
or we can move them parallel to the logical line, in which case the two
logical bits are adjacent to each other in a line and must be interleaved
(q0,q1,q2 with the next q0,q1,q2 just below it in the above figure).
}
\label{fig:2d_lattice}
\end{figure*}

If we put the circuit from Figure \ref{fig:circuit_opt} on the lattice in
Figure \ref{fig:2d_lattice}, we see that all the bits that interact in the
recovery circuit are already near one another in the lattice.  The only
additional complication we need to consider is
the difficulty of bringing three logical bits
near one another in order to do a logical operation.

\begin{figure}
\centering
\includegraphics{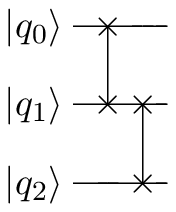}
\caption{A $SWAP_3$ gate, which is composed of two swaps on three bits.  Since this
gate only acts on three bits, we assume that its
error rate is at most $g$, the error of any 3-bit gate.}
\label{fig:swap3}
\end{figure}

To
operate on logical bits, we must interleave the bits, operate
locally, and then uninterleave.  But in which direction do we interleave?
There are two directions: parallel and perpendicular to the logical
bit line (see Figure \ref{fig:2d_lattice}).  Interleaving three
logical bits parallel to the logical line requires nine SWAP gates.  Interleaving
three logical bits perpendicular to the logic line requires $12$ SWAP
gates.  However, both schemes use at most six SWAPs on a
given logical bit.  If we combine two SWAPs into one three bit gate,
which we call $SWAP_3$,
depicted in Figure \ref{fig:swap3}, (and
then only count three bit gates in the threshold) we only use three $SWAP_3$
gates.

Thus, a full cycle now requires six additional gate operations: three to interleave,
three to uninterleave the bits\footnote{The error-recovery circuit in Figure
\ref{fig:circuit_opt} actually rotates the logical bit line, but as long as
all bits are recovered at the same time, this rotation is uniform throughout
the circuit and can be ignored}. Thus our total gate count is 14 if we ignore
bit initialization, and 16 if we include bit initialization.  As such the
threshold using only local operations in $2D$ becomes
$\rho_2 = 1/3{14 \choose 2}=1/273$ and $\rho_2 = 1/3{16 \choose 2}=1/360$
respectively.  Clearly, if we can initialize bits with error probability very much
lower than $1/273$, they can be ignored in the threshold calculation, and
we can assume that the gate error rate only needs to reach the larger
threshold, which is approximately $0.4\%$.

\subsection{A local 1-dimensional fault-tolerant system}
\label{sec:1dft}
\begin{figure}
\centering
\includegraphics{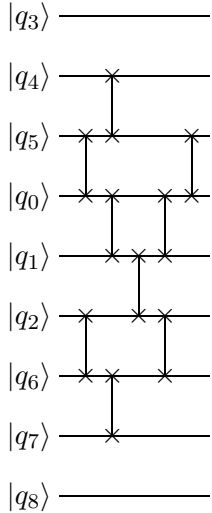}
\caption{Interleaving three codewords which are linearly adjacent.}
\label{fig:il_3}
\end{figure}

\begin{figure}
\scalebox{0.85}{
\includegraphics{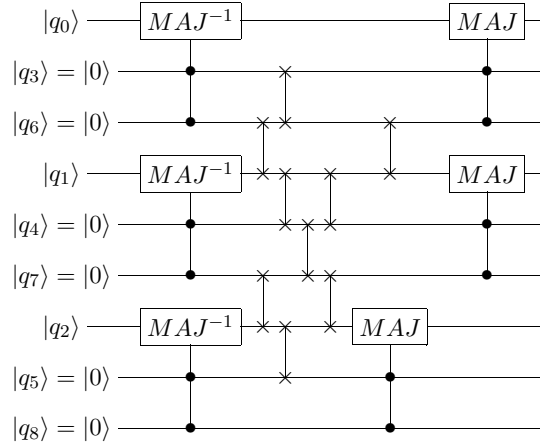}
}
\caption{
A fault-tolerant error-recovery circuit in one dimension with only
local gate operations.  This may be thought of as Figure \ref{fig:circuit_opt}
plus Figure \ref{fig:il_3} handling the the interleaving of the bits.
}
\label{fig:1d_circ}
\end{figure}
Figure \ref{fig:1d_circ} uses only nearest-neighbor operations on a 1D
array.  The error correction circuit requires six MAJ gates, nine SWAPs,
and six initializations.  Instead of counting nine SWAPs, we count four $SWAP_3$ gates
and one SWAP.  Instead of counting six initializations we count two 3-bit
initializations.  This gives a total of $11$ gates or $13$ gates, with or without
initialization, respectively.

We want to balance the number of SWAPs applied to each codeword so that no 
codeword is corrupted more than the other two.  As such, we interleave by
bringing the two outer codewords close to the middle codeword.
In order to interleave three logical bits $b0,b1,b2$, we move the
last bit in $b0$ just above the last bit in $b1$.  Then we move the second bit
in $b0$ just above the second bit in $b1$.  Next, we move the first bit in $b0$
just above the first bit in $b1$.  Subsequently, we do a similar operation on
$b2$.

Interleaving $b0$ and $b1$ requires $8+7+6$ SWAPs (8 for the last
bit, 7 for the second bit, 6 for the first bit).  Interleaving $b2$
requires $10 + 8 + 6$ SWAPs (10 for the first bit, 8 for the second, and 6 for
the last).  This gives a total of $45$ SWAPs; however, at most $24$ act on a
single bit.  If, instead of counting SWAPs, we count $SWAP_3$ gates, we have only
$12$ $SWAP_3$ gates acting on each codeword to interleave.

Thus, a full operation is $12$ $SWAP_3$ on each codeword to interleave, the
gate operation, which touches each of the three bits in each codeword, and
finally $12$ $SWAP_3$ gates to uninterleave, for a total of $27$ gates, in
addition to the error-recovery cycle.  The error-recovery cycle requires $13$
gates ($11$ if we neglect initialization), for a total of 40 gates.  This
yields a threshold of $\rho_1 = 1/3{40\choose 2}=1/2340$ (or $\rho_1 = 1/2109$ if bit
initialization is much more accurate than $\rho_1$).  Thus, we find that
$\rho_1$
is about an order of magnitude worse in the $1D$ case than it is in the $2D$
case.  In the next section, we will see how using a few levels of $2D$ at the
lowest level can recapture most of the advantage that $2D$ offers in the threshold.

\subsection{Concatenating different thresholds}
\label{sec:concat_thresh}
If a particular fault-tolerant scheme has a threshold $\rho_2$ and
the elementary gates have an error rate of $g$, Equation \ref{eq:g_k}
tells us that, after $k$ levels of
concatenation, the resulting error rate is less than:
\[
g_k = \rho_2\left( \frac{g}{\rho_2} \right)^{2^k}
\]

If, after $k$ levels of this scheme, we concatenate with $L-k$ levels
of a scheme with threshold $\rho_1$, we have:
\begin{eqnarray*}
g_L &=& \rho_1\left( \frac{g_k}{\rho_1} \right)^{2^{L-k}}\\
    &=& \rho_1\left( \frac{ \rho_2\left( \frac{g}{\rho_2} \right)^{2^k}
    }{\rho_1}\right)^{2^{L-k}}\\
    &=& \rho_1 \left(
    \frac{g}{\left(\frac{\rho_1}{\rho_2}\right)^{\frac{1}{2^k}} \rho_2 }
    \right)^{2^L}\\
    &=& \rho_1 \left(\frac{g}{\rho(k)}\right)^{2^L}
\end{eqnarray*}
with $\rho(k)=\rho_2\left(\frac{\rho_1}{\rho_2}\right)^{\frac{1}{2^k}}$.
Thus, if we use $k$ levels of a scheme with a lower threshold, we can
get most of the advantages of a lower threshold, even with a small,
finite number of concatenations.  Table \ref{tab:thresh_of_k} shows
that, after a few levels of $2D$ concatenation, the threshold approaches
the $2D$ case studied in Section \ref{sec:2dft}.  In particular, a linear
array nine bits wide has a threshold $60\%$ as large as the full $2D$ case, and
an array 27 bits wide has a threshold $77\%$ as large as the full $2D$ case.
This underscores the fact that most of the benefits of a $2D$ structure accrue
\emph{in the first few levels of concatenation}.
\begin{table}
\centering
\begin{tabular}{l|l|l}
k & Width & $\rho(k)/\rho_2$ \\
\hline\\
0 & 1 & 0.13\\
1 & 3 & 0.36\\
2 & 9 & 0.60\\
3 & 27 & 0.77\\
4 & 81 & 0.88\\
5 & 243 & 0.94
\end{tabular}
\caption{If we concatenate $k$ levels of 2D circuits with $L-k$
levels of 1D circuits, we see that the threshold rapidly approaches
the 2D case.  If we are allowed 27 lines rather than just one, we can
get a threshold which is only 23\% smaller than 2D case.}
\label{tab:thresh_of_k}
\end{table}

\section{Entropy dissipation}
\label{sec:entropy}
Reversible computing has been proposed as a method to reduce power
consumption of computing devices.  In some quantum systems, reversible
logic is all that is available, and irreversible devices must be
simulated from reversible ones (by discarding or resetting bits).
However, a Toffoli gate can simulate an irreversible NAND gate
by dissipating at most $3/2$ bits of entropy per cycle\footnote{
The value of $3/2$ bits is in fact optimal (assuming
equally likely inputs and  using only reversible logic),
and may be achieved using the $MAJ^{-1}$ gate.
}.
Similarly, due to the universality of NAND, we can use NAND gates to build a
functionally equivalent gate to Toffoli at a entropic expense of only a few
bits.
Hence, once our encoded gates
dissipate $3/2$ bits of entropy per operation, we can say that we have
actually used faulty, reversible gates to build fault-free \emph{irreversible}
logic.

It has been shown that if a reversible computer has errors, 
there must
be a supply of fresh zero bits in order to remove entropy from the
computer\cite{ABIN96}.  Here, we estimate how much entropy per gate must be
dissipated during fault-tolerant operation of a noisy reversible
computer.  We note that when n bits have $n\times H$ bits of entropy, it is not necessary
to replace them with $n$ zero-entropy bits; instead, reversible
cooling\cite{SV99,BMRVV02,FLMR04} schemes can ensure than we only need to
replace $n \times H$ of them with zero-entropy bits.  Thus, asymptotically, we
need to calculate the expected amount of entropy in the ancillary bits, and
that will correspond to the number of bit resets we will need in our system.

Landauer pointed out that where there is irreversibility in computing,
there must be heat dissipation\cite{Landauer61}.  Thus, by computing the
amount of entropy dissipated, we know that the heat dissipated is:
\[
\Delta E \ge k_b t \Delta H
\]
where $\Delta H$ is the amount of entropy dissipated, $k_b$ is Boltzmann's
constant, and $t$ is the temperature.

We represent the number of gates at level $L-1$ needed to simulate a gate at
level $L$ as $\tilde{G}$.
The value of $\tilde{G}$ 
will depend on the model we are working with, e.g. non-locally connected
vs. locally connected. 
Following the calculations in Section \ref{sec:circuit_blowup},
and due to subadditivity of entropy,
if $H_L$ is the entropy generated by one gate operation at level $L$, we
can see that:
\begin{eqnarray*}
H_L &\le& \tilde{G} H_{L-1}\\
     &=& \tilde{G}^{L-1} H_1
\end{eqnarray*}

Not every error is distinguishable, so assuming each error can be distinguished
provides an upper bound.  When a gate has an error, we assume it outputs
totally random bits; thus, with probability $1-g$, the output is correct, and
with probability $g$ the output is one of eight equally likely outputs.
Calculating the entropy of such a scenario yields $H(\frac{7g}{8}) +
\frac{7g}{8}\log 7$, and we find:
\begin{eqnarray*}
H_1 &\le& \tilde{G}\left(H(\frac{7g}{8}) + \frac{7g}{8}\log 7\right)\\
    &\le& \tilde{G}\left(2\sqrt{\frac{7g}{8}} + \frac{7g}{8}\log 7\right)\\
    &\le& \tilde{G}\left(2\sqrt{\frac{7}{8}} + \frac{7}{8}\log 7\right)\sqrt{g}
\end{eqnarray*}
If we define $\kappa = \left(2\sqrt{\frac{7}{8}} + \frac{7}{8}\log 7\right)$,
then we have:
\[
H_L \le \tilde{G}^L \kappa \sqrt{g}
\]

To obtain a lower bound, we recall that we assume independent errors.
From Figure \ref{fig:concat} we can see that errors in the recovery process
are independent of errors in different logical bits.
After each gate at level $L>0$, we do error-recovery, and this is
where the entropy is removed by means of bit resets.
If there are $E$ gates
in the recovery process, we know that:
\begin{eqnarray*}
H_L &\ge& 3E H_{L-1}\\
    &=& (3E)^{L-1} H_{1}
\end{eqnarray*}

We note that every gate touches at least one ancillary
bit.
With probability $g$ the physical gate fails.
When the gate fails, that bit will be flipped with
probability $1/2$; thus, the entropy of all the ancillary bits is at least:
\[
H_1 \ge H(g/2) \ge g
\]
So we have:
\[
H_L \ge (3E)^{L-1} g 
\]

Putting both the upper and lower bounds together:
\[
g (3E)^{L-1} \le H_L \le \tilde{G}^L \kappa \sqrt{g}
\]

If we want to have $O(1)$ bits of entropy per gate, by bounding
the left side of the above equation, we must have:
\begin{eqnarray*}
g (3E)^{L-1} &\le& 1\\
\log g + (L-1)\log 3E &\le& 0\\
L &\le& \frac{\log\frac{1}{g}}{\log 3E} + 1
\end{eqnarray*}
For example, if $g = 10^{-2}$, and $E=11$, we have $L \le 2.3$.

We see that the entropy-saving aspect of reversible computing is indeed
highly sensitive to error.
Both the upper and lower bounds of entropy per gate are exponential
in $L$ for fixed $g$.
At the same time, we see that, even if
there is some small finite error with $g \ll 1$, the entropic savings relative
to irreversible computing may be
obtained by using $O(\log\frac{1}{g})$ levels
of error correction.

\section{Conclusion}

We have given a method of producing fault-tolerant reversible circuits.
We also considered this problem in which only local communications are allowed,
which we believe will be very valuable for quantum computing systems that
need to perform some classical processing without having to resort to quantum
measurements\cite{VYWJBRMD00}.  We also note that the circuits and
threshold values presented here represent an \emph{lower} bound on the
threshold for reversible, fault-tolerant logic.  There may exist improved
schemes which could improve the threshold values; however, the circuits here
provide an existence proof. 

While we showed how to make fault-tolerant, reversible circuits, we also saw
that, when the error rate is near the threshold, there is considerable cost
(in bits, gates, and entropy) to the error correction procedure.  While it
was already known that, in principle, any noisy reversible computer must dissipate
entropy\cite{ABIN96}, our circuits provide a useful upper bound on how much
entropy must be released in the computing process with noisy reversible gates.

The authors would like to thank Michael Frank for many helpful comments,
and Thomas Szkopek for pointing out an error in an earlier draft.

\bibliographystyle{latex8}
\bibliography{revft}

\end{document}